# Quantitative analysis of the nanoscale intra-nuclear structural alterations in hippocampal cells in chronic alcoholism via transmission electron microscopy study

Peeyush Sahay[a], Pradeep Shukla[b], Hemendra M. Ghimire[a], Huda Almabadi[a],
Vibha Tripathi[a], Samarendra K. Mohanty[c], Radhakrishna Rao[b], Prabhakar Pradhan[a*]

[a]*Department of Physics, BioNanoPhotonics Laboratory, University of Memphis, Memphis, TN, USA*
[b]*Department of Physiology, University of Tennessee Health Science Center, Memphis, TN, USA*
[c]*Biophysics and Physiology Group, Department of Physics, University of Texas, Arlington, TX, USA*

*Corresponding author: ppradhan@memphis.edu*

## ABSTRACT

Chronic alcoholism is known to alter morphology of hippocampal, an important region of cognitive function in the brain. We performed quantification of nanoscale structural alterations in nuclei of hippocampal neuron cells due to chronic alcoholism, in mice model. Transmission electron microscopy images of the neuron cells were obtained and the degrees of structural alteration, in terms of mass density fluctuations, were determined using the recently developed light localization analysis technique. The results, obtained at the length scales ranging from 33 – 195 nm, show that the 4-week alcohol fed mice have higher degree of structural alteration in comparison to the control mice. The degree of structural alterations starts becoming significantly distinguishable ~100 nm sample length, which is the typical length scale of the building blocks of cells, such as DNA, RNA, etc. Different degrees of structural alterations at such length scales suggest possible structural rearrangement of chromatin inside the nuclei.

# INTRODUCTION

Hippocampus is an important part of human brain, which deals with brain functions, such as storing memories, emotional responses, navigation, etc.[1-3]. Morphological and chemical changes in hippocampal cells/tissues can significantly affect brain functions. External factor such as consumption of alcohol, including short term as well as long term, are known to bring such changes in the hippocampal cells/tissues, which has both, temporal and chronic adverse effects. One of the important cell categories in the hippocampal cells is the neural cell. Studies show that, as an immediate effect of the consumption of alcohol impedes the chemical signaling process in neural cells leading to alteration in brain's perception, which results in immediate abnormal behavioral acts. A significant body of evidence indicates that alcohol consumption causes damages to central nervous system[4]. However, a long-term consumption of alcohol can result in sustained structural damage to hippocampal cells/tissues potentially resulting in permanent morphological changes in the hippocampus volume[5-6]. Hippocampus being an important region of cognitive function is also a primary ethanol sensitive region in the brain[7]. Evidence indicates that the impairment of hippocampal integrity caused by alcohol consumption is caused by attenuation of neurogenesis[8-11]. Such structural change in hippocampus can result in neuropathological disorder, severely affecting day-to-day life activities of a person. Therefore, there has been a significant interest in studying the effect of alcohol on hippocampal cells.

In this work we are interested in analyzing the morphological changes (i.e. structural alterations) in the hippocampal cells as a result of long term exposure of alcohol, in alcoholic mice model. In particular, we are quantifying structural alterations in the nuclei of hippocampal neuron cells by employing a recently developed powerful light localization analysis approach, which has nano-scale level of detection sensitivity[12-13]. In this approach, the structural alteration is quantified in terms of the mass density fluctuations. In particular, the quantification measurement is performed in a single parameter called inverse participation ratio (IPR) values, which takes care all the structural heterogeneity of the system, and is termed as degrees of structural disorder $L_{sd}$. A detailed discussion on the technique and various terminologies associated with it are presented in the following sections.

The IPR approach is a very powerful technique, used in condensed matter physics, for analyzing heterogeneously disordered system via statistical analysis of spatially localized eigenfunctions of the light wave in the media[14-15]. The IPR value, for an eigenfunction $E$, is defined as IPR $=\int |E(r)|^4 dr$ (in the unit of inverse area). For an optical lattice system, if $E$ is its one of the eigenfunctions of the Hamiltonian, then the value of the IPR measures the degree of spatial localization of that eigenfunction. It has been shown that the strength/degree of light localization in a weakly optical disordered media increases linearly with the increase in the strength of the disorder of the system. Therefore, IPR value provides a measure of the degree of disorder inside an optical lattice system. The average value of the IPR at specific length scale of a uniform lattice is a fixed universal number (~2.45), but the value linearly increases with increase in the degree of disorder in the media. The quantum mechanical concept of IPR and its applications in characterizing heterogeneously disordered media, in a single parameter, can be seen elsewhere[14-21]. In TEM imaging, electron wave interaction with different parts of the sample modulates the intensity at the image plane. It has been shown that the value at the spatial location of the voxel (x, y) is $I_{TEM}(x, y) = I_{0\ TEM} + dI_{TEM}(x,y)$, which is related to the refractive index: $n(x,y) = n_0 + dn(x,y) = \rho_0 + \beta.\rho(x,y)$, where $I_0$ and $n_0$ are the respective mean values of TEM intensity and refractive index medium surrounding a scattering structure, $\rho(x,y)$ is local mass density of the biological media, and $\beta$ (≈0.185 for most biological molecules found in living cells) is the specific refraction increment. Therefore, intensity of fluctuation in TEM imaging is reflected in refractive index fluctuations by: $n(x,y)$ $\alpha\ M(x,y)\ \alpha\ I(x,y)$ [13, 22-23].

To obtain the eigenfunctions of the optical system, the Hamiltonian of the system is generated from the optical lattice. To do this, we use Anderson's disordered tight binding model (TBM) optical Hamiltonian, which has been well studied in condensed matter physics and proven to be a good model to quantify eigenfunction localization of the system of any type or geometry and disorder. A TBM Hamiltonian with $|i\rangle$ and $|j\rangle$ optical wave functions for respective $i^{th}$ and $j^{th}$ lattice sites becomes:

$$H = \sum_i \varepsilon_i |i\rangle\langle i| + t \sum_{\langle ij \rangle} (|i\rangle\langle j| + |j\rangle\langle i|), \qquad (1)$$

where $t$ is the overlap integral between $i^{th}$ and $j^{th}$ sites. To determine the effective optical potential of $i^{th}$ optical lattice, $\varepsilon_i$ for the voxel around the point $(x, y)$ can be written as $\varepsilon_i \propto dn(x,y)/n_o = dI_{TEM}/I_0$, since $dn(x,y) << n(x, y)$ and $dI_{TEM}(x, y) << I_0$. Now, we can define average value of *IPR* where the average is taken over all $N$ eigenfunctions of a sample size $LxL$ (i.e., $<IPR(L)>_{sample}$) in an optical lattices system as

$$\langle IPR(L) \rangle_{Sample} = \frac{1}{N} \sum_{i=1}^{N} \int_0^L \int_0^L E_i^4(x, y) dx dy, \qquad (2)$$

where $E_i$ is the $i^{th}$ eigenfunction of the Hamiltonian of optical lattice of size $L \times L$, and $dx = dy = $ a, is the length scale of each smallest pixel. Total number of eigenfunctions $N = (L/a)^2$. For the average *IPR*, the degree of structural disorder $L_{sd}$ can be written as

$$\langle IPR(L) \rangle_{Pixel} \propto L_{sd}. \qquad (3)$$

For the sake of simplicity, we will use the proportionality constant in the above equation as 1.

All work with animals was approved by the Animal Care and Use Committee of the University of Tennessee UT, Health Science Center, Memphis, TN, USA, in accordance with institutional and U.S. federal guidelines for animal experimentation. Adult female mice (C57BL6; 10-12 weeks) were caged with 2 in each group and administered with or without alcohol in their diet. One group of mice were fed a Lieber-DeCarli liquid diet (Dyets Inc., Bethlehem, PA) with f1-6% ethanol (0% for 2 days, 1% for 2 days, 2% for 2 days, 4% for one week, 5% for one week and 6% for one week), while the control group of animals in each group were pair-fed with an isocaloric diet adjusted with maltodextrin. Animals were euthanized and hippocampal samples were collected and fixed to suitable chemicals according to our probing instrument.

Biopsy samples obtained from hippocampal areas of mice were fixed more than two hours in 0.1 M Na cacodylate buffer (pH 7.2 to 7.4) with 2.5 percent glutaraldehyde and 2.5 percent paraformaldehyde. These fixed samples were then washed with several changes of 0.1 M cacodylate buffer (pH 7.2 to 7.4) to preserve structure. We post-fixed this sample with 2 percent osmium tetroxide in 0.1 M Na cacodylate buffer for 1-2 hours and rinsed it with several changes of 0.1 M cacodylate buffer (pH 7.2- 7.4). Following the standard protocol, samples were en bloc stained with aqueous UA and dehydrated through ethanol series every 15 minutes after rinsing with deionized water. For further dehydration, samples were embedded in polymer resin containing EPON to ensure sufficient stabilization for ultrathin sectioning. Thoroughly cleaned and concentrated samples prepared above were then sectioned with an ultra-microtome to the thickness of 70 nm so that the sample would be electron transparent and mechanically robust. Since the resultant specimen was sufficiently thin to allow the penetration of TEM electron beams, it was ready for imaging.

We used the classical Joel JEM-1200 TEM with its higher magnification (3000x) and working potential of 60 kV. Acquired wavelength by incident electron beam in TEM is less than nanometer, and we were able to get resolution down to fractions of a nanometer. TEM data were recorded in the form of images, i.e., 2D micrographs. Several (~12-15) neuron cells from hippocampal area were selected for the study via TEM imaging of the biopsy samples. The intra-nuclear disorder analysis was performed for the nucleus part of the cells after cropping them out from the whole cell image. A refractive index lattice system, '*optical lattice*', is created from the TEM image intensity data (as described in the Method). Subsequently, degrees of intra-nuclear disorder, $<L_{sd}>$, in terms of IPR values, were determined at different length scales for the each lattice system. In particular, sample length $L$ dependent average structural disorder $<L_{sd}(L)>$ values were determined at length scales ranging from ~ 33 – 200 nm. For every cell, the IPR value was calculated at each length scales, and were averaged over the whole cell to obtain a mean $<IPR(L)>_{pixel}$ value for the cell. Subsequently, an overall mean $<<IPR>_{pixel}>_{cell} \sim <L_{sd}>$ value for each category of cell was determined by taking the average of $<IPR(L)>_{pixel}$ values for all the cells of that category. Similarly, mean standard deviation in the $<L_{sd}>$ values, $(\sigma(L_{sd}))$, was also calculated for both, i.e., the alcoholic and non-alcoholic categories of the cells. The results are presented in the following section.

**RESULT S AND DISCUSSION**

Figures 1(a) and 1(a') show the actual TEM images of nucleus of sample neuron cells of non-alcoholic and alcoholic mice, respectively. Conversely, figures 1(b) and 1(b') are their respective $<L_{sd}(L)>$ images computed at length $L=130$ nm. The color map from blue to red represents increasing order of degree of disorder, i.e., the $<L_{sd}>$ values, inside the nucleus. As it can be seen that, the neuron cell nucleus of the alcoholic mice has more hot spots compared to the non-alcoholic mice suggesting that there is higher degree of disorder, i.e., mass density fluctuation inside them. A better angle view of the disorder map for both the cells has been represented in fig 1(c) and 1(c'). The regions of higher peaks are the areas where the fluctuation in mass density, i.e., $d\rho(x,y)$, is higher compared to average mass density of the cell. It is to be noted that the term $d\rho(x,y)$ represents the strength of the mass density fluctuation, therefore, the region of high peaks equally indicates about either a lower or higher mass density, in comparison to the average mass density of the cell nucleus, in that area.

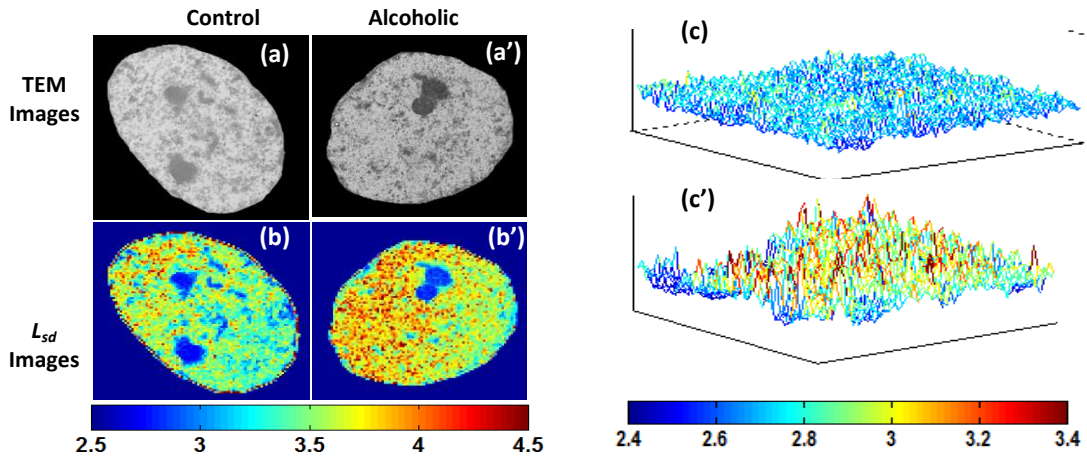

**Figure 1.** (a) and (a') are representative TEM images of nucleus from hippocampus of normal and chronic alcoholic mouse. (b) and (b') are corresponding $<L_{sd}>$-sample images for pixel size L×L=130 nm x 130 nm (TEM resolution 6.52 nm). (c) and (c') are respective 3-dimensional views for the $L_{sd}(x,y)$ at same length..

Fig. 2 shows the average disorder, $<L_{sd}>$, measured at sample length 163 nm. For the mice treated regularly with alcohol, the $<L_{sd}>$ value at 163 nm length scale was determined to be 4.05, where the $<L_{sd}>$ value for the non-alcoholic (control) mice was found to be 3.77. The *student's t* test obtained a p-value of 0.0015, suggesting that the average degrees of disorder determined in both the cases are significantly different. The results for standard deviation (std) in the disorder, $\sigma(L_{sd})$, at the same sample length, i.e., $L \times L = 163\ nm \times 163\ nm$, is shown in Fig. 2(b). The results clearly exhibit that the alcoholic mice cells have considerably higher standard deviation in their $\sigma(L_{sd})$ values. This suggests that the alcoholic mice have higher mass density fluctuations in their neuron cell nucleus in comparison to the control mice. Subsequently, we also compared the $<L_{sd}(L)>$ values in alcoholic and control mice at different sample lengths $L$. The Fig. 3(a) shows the plot of mean $<L_{sd}(L)>$ vs. $L$ for different sample lengths ranging from 33 nm to 195 nm. As it can be seen in Fig. 3(a), when the sample length $L$ is increased from $L = 163\ nm$, the mean $<L_{sd}(L)>$ values also increases, as well as the differences between the $<L_{sd}(L)>$ values of alcoholic mice and the non-alcoholic mice. Further in Fig. 3(a), it can be seen that $<L_{sd}(L)>$ values decreases with the decrease in the sample $L$; accordingly the difference between the $<L_{sd}(L)>$ values for alcoholic mice and the control mice decreases as well. It is worth pointing out here that the visible difference is prominent around the length scale ~100 nm, which is the typical dimension of the building blocks of the cell nuclei, such as DNA, RNA. etc.

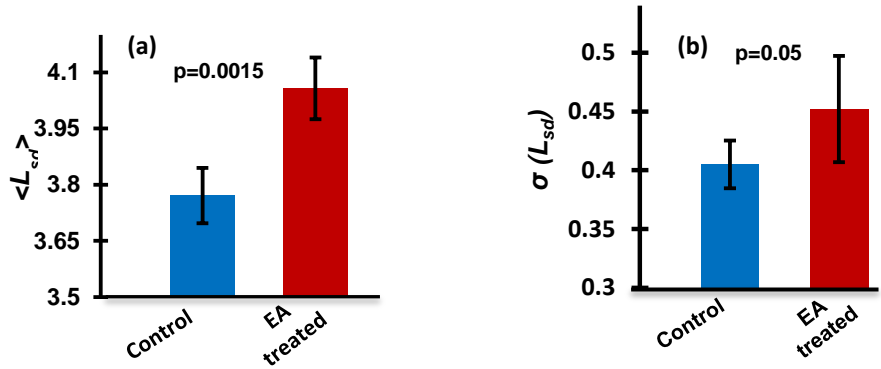

**Figure 2.(a).** Average $L_{sd}$ bar plot at pixel size $L \times L = 163\ nm \times 163\ nm$ and ensemble averaging for (i) hippocampus neural cell nuclei of control mice (control) and (ii) chronic alcoholic mice (EA treated). **(b)** Corresponding standard deviation $\sigma(L_{sd})$ at $L \times L = 163\ nm \times 163\ nm$.

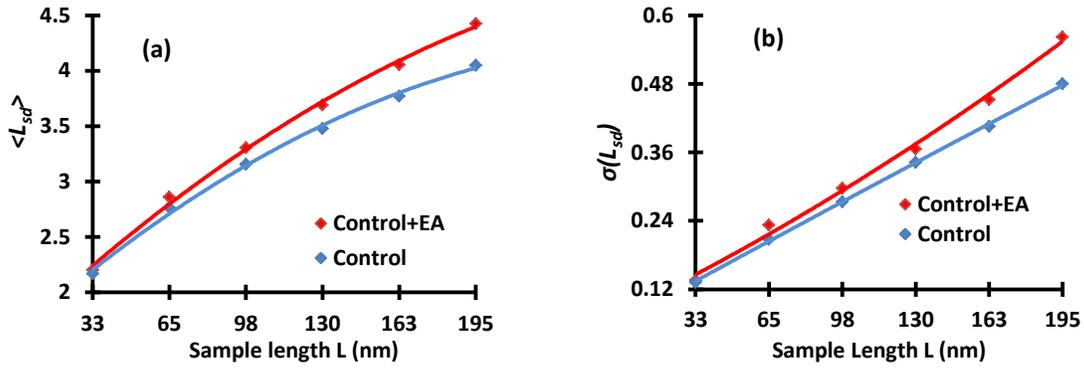

**Figure 3(a).** Graphical representation of ensemble-averaged values of $<L_{sd}(L)>$ versus $L$ (in nm) plots for (i) nuclei of normal mice (ii) nuclei of alcoholic mice. **(b)** Corresponding standard deviation $\sigma(L_{sd})$ versus $L$ (in nm) plots for nuclei of normal mice and alcoholic mice.

It should be noted that the degree of disorder, $<L_{sd}(L)>$, which is measured in terms of IPR numbers and calculated at a particular sample length $L$, basically measures the degree of spatial inhomogeneity in the refractive index distribution (and, thus in the mass density distribution) inside the sample area size $L \times L$. Therefore, it is reasonable to expect the $<L_{sd}(L)>$ values will increase with large $L$, as the strength of inhomogeneity increases with increase in $L$ till a saturation value of the IPR is reached at very large $L$[12-13]. Expectedly, therefore, at higher length scales, the $<L_{sd}(L)>$ values will increase, and so will the difference between the $<L_{sd}(L)>$ values of two disordered system. A similar trend can be seen for the standard deviation (std) in $<L_{sd}(L)>$ values, $\sigma(L_{sd})$, for both the cases, Fig. 3(b). It is interesting to note that the present system (i.e., the biological sample) exhibits perfect example of a weakly disordered system where the mean and the standard deviation of disorder parameter increases with the increase in the sample size till a saturation is reached, implying that the nanoscale structural disorder is increasing with alcoholism. Such systems are well characterized by the IPR method, based on mesoscopic physics.

In terms of mass density fluctuations, a higher $<L_{sd}(L)>$ value at a sample length $L$ indicates a higher mass density fluctuations, or rearrangement, in spatial mass density distribution in the sample area $L \times L$. Therefore the present results suggest that the alcoholic mice have higher mass density fluctuation inside the nuclei of their hippocampal neural cells compared to the non-alcoholic (control) mice. Similarly, a higher standard deviation in $L_{sd}$, $\sigma(L_{sd}(L))$, values in alcoholic mice suggest a more spread in mass density fluctuations in them compared to the non-alcoholic mice. It is important to point out here that, the nucleus inside a cell is densely packed with the chromatin coils. Therefore, the results obtained in this study indicate about a possible rearrangement in the chromatin structure. Changes in protein expression profiles have been recorded in the hippocampus of alcoholic subjects [24]. Additionally, it is also interesting to point out here is that a typical diameter of chromatin coil is ~30 nm. Our measurement shows not a significant difference in the $<L_{sd}(L)>$ values at this length scales in the two cases, however aroung three times of this length scale. These results invite more research in this direction in order to elucidate the actual effects of chronic alcoholism on brain hippocampal neuron cells.

In summary, we studied quantification of nanoscale structural alteration, in terms of mass density fluctuations, in hippocampal neuron cells as an effect of chronic alcohol consumption. Using the powerful IPR analysis approach, the study was performed on TEM images of hippocampal neuron cells of the non-alcoholic (control) and alcoholic mice. The structural disorders were evaluated inside the nucleus of the hippocampal neuron cells at various length scales ranging from 33 – 195 nm. The results show higher degree of structural alterations in alcoholic mice in comparison to the control mice, which suggests about possible rearrangement of chromatin structure inside the nucleus.

**ACKNOWLEDGEMENT:**

This work was supported by NIH R01 EB003682 and UofM (PPradhan), and NIH AA12307 and DK55532 (RKRao).

**REFERENCES:**

1. Andersen, P. *et al*., eds., *The Hippocampus Book* (Oxford University Press, 2006),

2. O'Keefe, J. & Nadel, L. *The Hippocampus as a Cognitive Map* (Oxford University Press, 1978)

3. Burgess, N., Maguire, E.A., & O'Keefe, J. The Human Hippocampus and Spatial and Episodic Memory. *Neuron* **35**, 625–641 (2002).

4. Walker D.W., Barnes D.E., Zornetzer S.F., Hunter B.E., & Kubanis P. Neuronal loss in hippocampus induced by prolonged ethanol consumption in rats. *Science* **209**, 711–713 (1980).

5. Riley, J.N., & Walker, D.W. Morphological Alterations in Hippocampus after Long-Term Alcohol Consumption in Mice. *Science* **201,** 646-648 (1978).

6. McMullen, P.A., Saint-Cyr, J.A., & Carlen, P.L. Morphological Alterations in Rat CA1 Hippocampal Pyramidal Cell Dendrites Resulting from Chronic Ethanol Consumption and Withdrawal. *The Journal of Comparative Neurology* **225**, 111–118 (1984).

7. Vilpoux, C., Warnault, V., Pierrefiche, O., Daoust, M., & Naassila, M. Ethanol-sensitive brain regions in rat and mouse: a cartographic review, using immediate early gene expression. *Alcohol Clin Exp Res* **33**, 945–969 (2009).

8. Collins M.A., Corse T.D., & Neafsey E.J. Neuronal degeneration in rat cerebrocortical and olfactory regions during subchronic "binge" intoxication with ethanol: possible explanation for olfactory deficits in alcoholics. *Alcohol. Clin. Exp. Res.* **20**, 284–292 (1996).

9. Bengochea O., & Gonzalo L.M., Effect of chronic alcoholism on the human hippocampus. *Histol. Histopathol.* **5**, 349–357 (1990).

10. Geil, C.R., Hayes, D.M., McClain, J.A., Liput, D.J., Marshall, S.A., Chen, K.Y., & Nixon, K. Alcohol and adult hippocampal neurogenesis: Promiscuous drug, wanton effects. *Prog Neuropsychopharmacol. Biol. Psychiatry* **3**, 103–113 (2014).

11. Nixon K, Kim D.H., Potts E.N., He J., & Crews F.T. Distinct cell proliferation events during abstinence after alcohol dependence: microglia proliferation precedes neurogenesis. *Neurobiol Dis.* **31**, 218–29 (2008).

12. Pradhan, P., Damania, D., Joshi, H.M., Turzhitsky, V., Subramanian, H., Roy, H.K., Taflove, A., Dravid, V.P., Backman, V. Quantification of nanoscale density fluctuations using electron microscopy: Light-localization properties of biological cells. *Appl. Phys. Lett.* **97,** 243704 (2010).

13. Pradhan, P., Damania, D., Joshi, H.M., Turzhitsky, V., Subramanian, H., Roy, H.K., Taflove, A., Dravid, V.P., Backman, V. Quantification of nanoscale density fluctuations by electron microscopy: probing cellular alteration in early carcinogenesis. *Phys. biol.* **8**.2, 243704 (2011).

14. Pradhan, P. & Sridhar, S. Correlations due to Localization in Quantum Eigenfunctions of Disordered Microwave Cavities. *Phys. Rev. Lett.* **85**, 2360–2363 (2000).

15. Pradhan, P. & Sridhar, S. From chaos to disorder: Statistics of the eigenfunctions of microwave cavities. *Pramana* **58**, 333–341 (2002).

16. Lee, P.A. & Ramakrishnan, T.V. Disordered electronic systems. *Rev. Mod. Phys.* **57**, 287–337 (1985).

17. Abrahams, E., Anderson, P.W., Licciardello, D.C., & Ramakrishnan, T.V. Scaling theory of localization—Absence of quantum diffusion in two dimensions. *Phys. Rev. Lett.* **42**, 673–676 (1979).

18. Kramer, B & Mackinnon, A. Localization—theory and experiment. *Rep. Prog. Phys.* **56**, 1469–1564 (1993).

19. Prigodin, V.N. & Altshuler, B.L. Long-range spatial correlations of eigenfunctions in quantum disordered systems. *Phys. Rev. Lett.* **80**:9, 1944 (1998).

20. Murphy, N.C., Wortis, R. & Atkinson, W.A. Generalized inverse participation ratio as a possible measure of localization ratio as a possible measure of localization for interacting systems. *Phys. Rev.* B **83**:18, 184206 (2011).

21. Jackson, S.R., Khoo, T.J., & Strauch, F.W. Quantum walks on tree with disorder: Decay, diffusion and localization. *Phys. Rev.* A, **86**:2, 022335 (2012).

22. Schmitt, J.M., & Kumar, G., Turbulent nature of refractive-index variations in biological tissue. *Opt. Lett.* **21**, 1310-1312 (1996).

23. Barer, R., Ross, K.F.A., and Tkaczyk, S. Refractometry of Living Cells. *Nature* **171**, 720–724 (1953).

24. Matsumoto, H & Matsumoto, I., Alcoholism: Protein Expression Profiles in a Human Hippocampal Model. *Expert Review of Proteomics* **5**:2, 321–231, (2008).